\documentclass[10pt,journal,epsfig]{IEEEtran}
\usepackage{graphicx}
\usepackage{subfigure,color}
\usepackage{amssymb}
\usepackage{cite,comment}
\usepackage{amsmath}
\usepackage{bm,comment,setspace}
\usepackage{algorithm}
\usepackage{algpseudocode}
\makeatletter

\newcommand{\Rmnum}[1]{\expandafter\@slowromancap\romannumeral #1@}

\makeatother

\newtheorem{remark}{\underline{Remark}}

\newtheorem{lemma}{\underline{Lemma}}

\begin{document}
\title{UAV-Sensing-Assisted Cellular Interference Coordination: A Cognitive Radio Approach}
\author{Weidong Mei and Rui Zhang, \IEEEmembership{Fellow, IEEE}
\thanks{\footnotesize{W. Mei is with the NUS Graduate School for Integrative Sciences and Engineering, National University of Singapore, Singapore 119077, and also with the Department of Electrical and Computer Engineering, National University of Singapore, Singapore 117583 (e-mail: wmei@u.nus.edu).}}
\thanks{\footnotesize{R. Zhang is with the Department of Electrical and Computer Engineering, National University of Singapore, Singapore 117583 (e-mail: elezhang@nus.edu.sg).}}}
\maketitle

\begin{abstract}
Aerial-ground interference mitigation has been deemed as the main challenge in realizing cellular-connected unmanned aerial vehicle (UAV) communications. Due to the line-of-sight (LoS)-dominant air-ground channels, the UAV generates/suffers much stronger interference to/from cellular base stations (BSs) over a much larger region in its uplink/downlink communication, as compared to the terrestrial users. As a result, conventional inter-cell interference coordination (ICIC) techniques catered for terrestrial networks become ineffective in mitigating the more severe UAV-induced interference. To deal with this new challenge, this letter introduces a cognitive radio based solution by treating the UAV and terrestrial users as secondary and primary users in the network, respectively. In particular, the LoS channels with terrestrial BSs/users endow the UAV with a powerful spectrum sensing capability for detecting the terrestrial signals over a much larger region than its serving BS. By exploiting this unique feature, we propose a new UAV-sensing-assisted ICIC design for both the UAV downlink and uplink communications. Specifically, the UAV senses its received interference and the transmissions of terrestrial users in the downlink and uplink, respectively, over the resource blocks (RBs) available at its serving BS to assist its RB allocation to the UAV for avoiding the interference with co-channel terrestrial communications. Numerical results demonstrate that the proposed UAV-assisted ICIC outperforms the conventional terrestrial ICIC by engaging the neighboring BSs for cooperation only.
\end{abstract}

\section{Introduction}
The popularity of unmanned aerial vehicles (UAVs) (a.k.a. drones) has skyrocketed over the last decade. This is mainly attributed to the fact that they have been made more easily accessible to the civilian users than ever before, thanks to their steadily decreasing cost and improved portability. In addition, this is also driven by the intense demands for them in various new applications, such as public show, package delivery, and communication platform\cite{zeng2019accessing}. However, the expanding UAV market, in turn, places more stringent requirements on the performance of UAV communications in these applications, in a need to support their large-scale deployment in the future. Unfortunately, existing UAV-ground communication, primarily relying on point-to-point link within the visual line-of-sight (LoS) range, can barely fulfill this goal\cite{zeng2019accessing}.

\begin{figure}[!t]
\centering
\includegraphics[width=3.6in]{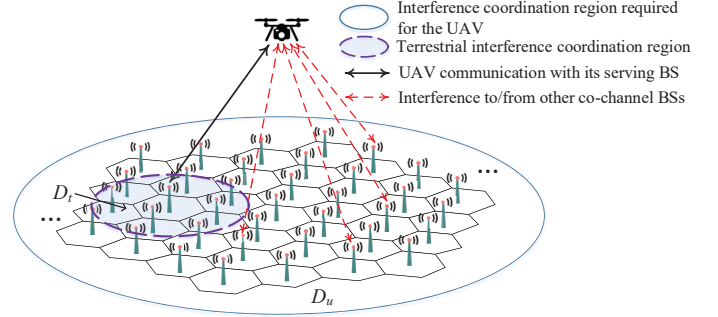}
\DeclareGraphicsExtensions.
\caption{Cellular-connected UAV communication subjected to aerial-ground interference.}\label{ccuc}
\vspace{-15pt}
\end{figure}
Recently, \emph{cellular-connected UAV} has emerged as an appealing solution to the above problem, by reusing the existing cellular base stations (BSs) and spectrum to serve UAVs as new aerial user equipments (UEs)\cite{zeng2019cellular,lin2019mobile}. In fact, preliminary field trials\cite{3GPP36777} and performance analysis\cite{yin2019uplink} have demonstrated that the current fourth-generation (4G) long-term evolution (LTE) network is able to meet the basic requirements of UAV communications. However, more severe aerial-ground interference than terrestrial communications has also been reported in \cite{3GPP36777} due to the LoS-dominant channels between the UAV and terrestrial BSs. As shown in Fig.\,\ref{ccuc}, in the downlink, the UAV suffers more severe inter-cell interference (ICI) from a much larger number of non-associated co-channel BSs in a much wider area $D_u$ as compared to terrestrial UEs, which can result in very poor achievable rate of the UAV. Whereas in the uplink, the UAV imposes severe ICI to a large number of co-channel terrestrial UE communications in $D_u$, thus causing significant rate loss of the terrestrial uplink transmissions. Although various ICI coordination (ICIC) techniques have been widely applied in terrestrial networks\cite{hamza2013survey}, they are generally applicable to a local region (see $D_t$ in Fig.\,\ref{ccuc}), which become ineffective to mitigate the aggravated aerial-ground interference in a much larger region ($D_u$). To improve over the conventional ICIC, several enhanced interference mitigation techniques for UAV have been proposed recently in \cite{liu2018multi,cellular2018mei,mei2019uplink,pang2019uplink} by engaging all BSs in $D_u$ for ICIC. However, they usually require high implementation complexity and low-latency backhaul links among the cooperating BSs, which may not be practically affordable.

In this letter, we propose a new, effective, and yet low-complexity ICIC design for cellular-connected UAVs, by leveraging the celebrated cognitive radio approach\cite{zhang2010dynamic,kang2009sensing} for achieving UAV sensing enabled opportunistic spectrum sharing. Specifically, we regard the UAVs as secondary UEs in the cellular network, which share the same spectrum as the terrestrial UEs (regarded as primary UEs), while effectively mitigating the interference to/from them in the UAV uplink/downlink communication\footnote{Note that a straightforward solution to avoid the aerial-ground interference is by reserving a fraction of the cellular spectrum for the exclusive use of the UAV UEs, regardless of whether they are present in any region. However, due to the heavy frequency reuse in the existing cellular network, this will result in low spectrum efficiency of the terrestrial communications as the number of reserved channels increases with that of the UAVs.}. In particular, for cognitive (secondary) UAV UEs, their LoS-dominant channels with the terrestrial BSs and UEs in fact endow them with a powerful spectrum sensing capability for detecting terrestrial signals over a much larger region than their serving BSs. By exploiting this new advantage, the UAV can detect the transmissions of terrestrial BSs/UEs in a wide area of $D_u$, thereby assisting its serving BS in resource block (RB) allocation to avoid the strong ICI in its downlink/uplink communication. Note that this requires only local cooperation between the UAV and its serving BS, but significantly enlarges the interference mitigation region as compared to the conventional ICIC (see $D_u$ versus $D_t$ in Fig.\,\ref{ccuc}).

Motivated by the above, in this letter we propose a new UAV-sensing-assisted ICIC design for both UAV downlink and uplink communications. Specifically, in the downlink, the UAV senses its received interference over the available RBs at its serving BS to help select low-interference RBs for maximizing its achievable rate. Whereas in the uplink, the UAV senses the terrestrial UEs' uplink transmissions in each available RB to predict its worst-case interfering channels to the co-channel BSs, based on which it helps its serving BS select the RBs with low sensed power and allocates its transmit power over them subject to the worst-case interference power constraints for protecting the terrestrial uplink transmissions. Numerical results demonstrate that the proposed UAV-sensing-assisted ICIC yields significant performance gains over the conventional ICIC without employing the UAV sensing.

\section{System Model}
As shown in Fig.\,\ref{ccuc}, we consider both the downlink and uplink UAV communications in a given region $D_u$ of the cellular network, where the ground BSs serve one single UAV (secondary) UE and a set of terrestrial (primary) UEs\footnote{Note that this work can be easily extended to the case with multiple UAVs in the region by applying the essential ideas in the later proposed UAV-sensing-assisted protocols, by e.g., assigning one or more UAVs into different sets of $N$ RBs.}. Each BS is assumed to be at the center of its located cell. For the BSs outside $D_u$, we assume that their interference to/from the UAV is attenuated to the level below the background noise, and thus can be ignored. Consequently, we only need to consider the interference to/from the BSs in $D_u$ for the UAV uplink/downlink communication.

We consider one snapshot of the network with given terrestrial UE and UAV locations, while the proposed designs can be similarly applied to other time instants.  We consider a total number of $N$ RBs, denoted by the set ${\cal N}\triangleq\{1,2,\cdots,N\}$. Due to frequency reuse in the cellular network, each of these RBs may have already been occupied by some terrestrial UEs in $D_u$. According to its practical rate demand, the UAV is assumed to request $N_d \le N$ and $N_u \le N$ RBs from $\cal N$ for its downlink and uplink communications, respectively. It is worth noting that $N_u \ge N_d$ usually holds in practice. This is because the UAV uplink is mainly catered for payload communication (e.g., video streaming), which requires much higher transmission rate than its control and non-payload communication (CNPC) in the downlink\cite{zeng2019cellular}. Centered at the UAV's horizontal location projected on the ground, we assume that there are in total $J$ BSs located in $D_u$ and denote them by ${\cal J}\triangleq\{1,2,\cdots,J\}$. Accordingly, we define a set ${\cal J}(n) \subseteq \cal J$ for each RB $n \in {\cal N}$, in which $j \in {\cal J}(n)$ if RB $n$ is occupied (already used) by a terrestrial UE in cell $j$.

\subsection{Cellular Downlink with New UAV UE Added}
Let $F_j(n)$ denote the downlink channel power gain from BS $j$ to the UAV in RB $n$, which in general depends on the BS/UAV locations, antenna gains, LoS/non-LoS (NLoS) channel path-loss and small-scale fading models\cite{zeng2019accessing}. Note that in the current 4G LTE network, the ground BS antennas are usually tilted downwards for mitigating the terrestrial ICI\cite{3GPP36777}. As such, we assume that each BS employs an antenna array with fixed directional gain pattern, while the UAV and all terrestrial UEs are assumed to be equipped with a single omnidirectional antenna for simplicity. It is also worth noting that for high-altitude UAVs (e.g., over 100 meters (m) above the ground), the LoS path-loss would dominate over the other NLoS components\cite{3GPP36777} and result in approximately frequency-flat UAV-BS channels over the spectrum. We thus consider that the UAV is associated with a BS $j_u \in \cal J$, which has the largest LoS channel gain (or smallest path-loss) with the UAV among all the BSs in $D_u$. Denote by $P_j(n) \ge 0$ the transmit power of BS $j \in \cal J$ in RB $n \in \cal N$. Then, the UAV's achievable rate in each RB $n \in \cal N$ is
\begin{equation}\label{sinr1}
R_{\text{DL}}(n) = \log_2\left(1+\frac{P_{j_u}(n)F_{j_u}(n)}{\sigma^2 + I_{\text{DL}}(n)}\right),
\end{equation}
in bits per second per Hertz (bps/Hz), where $\sigma^2$ denotes the Gaussian noise power at the UAV receiver, and $I_{\text{DL}}(n) = \sum\nolimits_{i \in {\cal J}(n)} P_i(n)F_i(n)$ denotes the total terrestrial ICI (assumed to be Gaussian) power at the UAV in RB $n$. Let ${\cal N}_{\text{DL}}$ denote the set of RBs assigned to the UAV by its serving BS $j_u$ in the downlink, with $\lvert {\cal N}_{\text{DL}} \rvert = N_d$. Then, the UAV's achievable sum-rate over ${\cal N}_{\text{DL}}$ is
\begin{equation}\label{sum1}
R_{\text{DL}} = \sum\nolimits_{n \in {\cal N}_{\text{DL}}} R_{\text{DL}}(n).
\end{equation}

It is noted from (\ref{sinr1}) and (\ref{sum1}) that $R_{\text{DL}}$ of the UAV depends on the RB allocations ${\cal N}_{\text{DL}}$ at its serving BS. Obviously, $R_{\text{DL}}$ can be improved if the UAV is assigned with RBs with lower interference powers or $I_{\text{DL}}(n)$'s. Conversely, $R_{\text{DL}}$ may become extremely low if each RB $n \in {\cal N}_{\text{DL}}$ has a large $I_{\text{DL}}(n)$. However, in practice, $I_{\text{DL}}(n)$'s cannot be obtained by the UAV's serving BS directly.

\subsection{Cellular Uplink with New UAV UE Added}\label{uplink}
Let $G_j(n)$ denote the uplink channel power gain from the UAV to BS $j$ in RB $n$. Since the uplink and downlink channels share the same path-loss, we consider that the UAV is served by the same BS $j_u$ as in the downlink. If the UAV transmits with power $p_u(n)$ in RB $n$, then its achievable rate in this RB is
\begin{equation}\label{sum2}
R_{\text{UL}}(n) = {\log_2}\left(1 + \frac{p_u(n)G_{j_u}(n)}{\sigma^2}\right),
\end{equation}
where $\sigma^2$ denotes the receiver Gaussian noise power at the UAV's serving BS\footnote{Here it is assumed that the terrestrial ICI has been mitigated to the level below the background noise by applying the conventional ICIC (see Section \ref{conv} for details).}. Let ${\cal N}_{\text{UL}}$ denote the set of RBs assigned to the UAV by its serving BS $j_u$ in the uplink, with $\lvert {\cal N}_{\text{UL}} \rvert = N_u$. Then, the UAV's achievable sum-rate over ${\cal N}_{\text{UL}}$ is
\begin{equation}\label{sum3}
R_{\text{UL}} = \sum\nolimits_{n \in {\cal N}_{\text{UL}}} R_{\text{UL}}(n).
\end{equation}

Meanwhile, the terrestrial UEs transmitting in ${\cal N}_{\text{UL}}$ in the uplink would suffer the co-channel interference from the UAV at their respective serving BSs. Accordingly, the maximum interference power imposed by the UAV to any co-channel terrestrial UE transmission over ${\cal N}_{\text{UL}}$ is expressed as
\begin{equation}\label{sum4}
I_{\text{UL}} =\mathop {\max}\limits_{n \in {\cal N}_{\text{UL}}, j \in {\cal J}(n)} p_u(n)G_j(n).
\end{equation}

It follows from (\ref{sum3}) and (\ref{sum4}) that there is a fundamental tradeoff between maximizing $R_{\text{UL}}$ and minimizing $I_{\text{UL}}$. For example, if the UAV increases its transmit power $p_u(n)$ in any RB $n$ to enhance the former, the latter may also increase, thus imposing stronger worst-case interference to the co-channel terrestrial UEs. Nonetheless, a proper RB allocation ${\cal N}_{\text{UL}}$ could help resolve the above tradeoff. For example, if the UAV is assigned with RBs with low interference channel power $G_j(n)$'s while maintaining high channel power $G_{j_u}(n)$'s with its serving BS, $I_{\text{UL}}$ can be greatly reduced without compromising $R_{\text{UL}}$. However, in practice, it is difficult for the UAV's serving BS to obtain $G_j(n), \forall j \in {\cal J}(n), n \in \cal N$ directly.

\section{Conventional ICIC}\label{conv}
In this section, we introduce the conventional ICIC designed for mitigating the terrestrial ICI in the cellular network, which will serve as a baseline for comparison with our proposed new UAV-sensing-assisted ICIC in Section \ref{new}.

In the conventional ICIC, each BS checks the availability of an RB in its first $q$ tiers ($q \ge 1$) of neighboring BSs (see the region $D_t$ in Fig.\,\ref{ccuc}) before assigning it to a new terrestrial UE. Let $N_j(q)$ denote the set of the first $q$-tier neighboring BSs of BS $j \in {\cal J}$ including itself. If an RB has been assigned to a terrestrial UE in $N_j(q)$, BS $j$ cannot assign this RB to any new terrestrial UE. By this means, BS $j$ and its served UEs will not cause any interference to all the other cells in $N_j(q)$ in the downlink and uplink, respectively. Note that when $q$ is sufficiently large, the terrestrial ICI would become negligible, thanks to the significant path-loss, shadowing, and multi-path fading of practical terrestrial channels. Next, we apply the above conventional ICIC to the new UAV UE in the downlink and uplink, respectively.

\subsection{Downlink RB Allocation}\label{conv1}
In the downlink, the UAV's serving BS $j_u$ first determines the set of available RBs based on the criterion introduced above. In particular, an RB $n \in \cal N$ is available at BS $j_u$ if $N_{j_u}(q) \cap {\cal J}(n) = \emptyset$, i.e., RB $n$ has not been assigned to any terrestrial UE in $N_{j_u}(q)$ yet. Let $\Omega_d$ denote the set of all available RBs at BS $j_u$, which is assumed to be sufficiently large such that $\lvert \Omega_d \rvert \gg N_d$. Next, BS $j_u$ randomly assigns $N_d$ RBs from $\Omega_d$, denoted as $\Omega_d'$, to the UAV and allocates its transmit powers in these RBs. We consider the following peak power constraints at the UAV's serving BS $j_u$, i.e., $P_{j_u}(n) \le P_{\text{DL}}, \forall n \in \cal N$, with $P_{\text{DL}}$ denoting the maximum allowable transmit power per RB for BS $j_u$. To maximize the UAV's achievable sum-rate, BS $j_u$ transmits with the peak power $P_{\text{DL}}$ in each RB $n \in \Omega_d'$. Thus, we have ${\cal N}_{\text{DL}}=\Omega_d'$ and $P_{j_u}(n)=P_{\text{DL}}, \forall n \in \Omega_d'$ in (\ref{sum1}) for the conventional ICIC.

Notice that with the conventional ICIC, the UAV is free of terrestrial interference from the BSs in $N_{j_u}(q)$ only. However, the RB allocation for the UAV neglects the interference from other BSs outside $N_{j_u}(q)$ but still in $D_u$ (see Fig.\,\ref{ccuc}), which may still cause strong interference to the UAV due to their LoS-dominant channels with the UAV. As a result, the UAV's downlink achievable rate will be severely limited.

\subsection{Uplink RB Allocation}\label{conv2}
Similar to the downlink, in the uplink, BS $j_u$ first determines the set of available RBs, denoted by $\Omega_u$ with $\lvert \Omega_u \rvert \gg N_u$. Then, BS $j_u$ randomly assigns $N_u$ RBs from $\Omega_u$ to the UAV, denoted as $\Omega_u'$. Consider the following peak power constraints at the UAV, i.e., $p_u(n) \le P_{\text{UL}}, \forall n \in \Omega_u'$. To maximize its achievable sum-rate, the UAV transmits with the peak power $P_{\text{UL}}$ in each RB $n \in \Omega_u'$. As such, the conventional ICIC leads to ${\cal N}_{\text{UL}}=\Omega_u'$ and $p_u(n)=P_{\text{UL}}, \forall n \in \Omega_u'$ in both (\ref{sum3}) and (\ref{sum4}).

Although the above RB allocation for the UAV by the conventional ICIC ensures that the UAV will not cause any interference to the uplink transmissions of terrestrial UEs in $N_{j_u}(q)$, it overlooks the UAV's potential interference to other co-channel terrestrial BSs in $D_u$. As a result, $I_{\text{UL}}$ in (\ref{sum4}) can be practically high due to the UAV's LoS-dominant channels with such BSs and the terrestrial uplink communication performance will be severely affected.

\section{UAV-Sensing-Assisted ICIC}\label{new}
It is worth noting that the fundamental limitation of the conventional ICIC for UAV RB allocation in both the downlink and uplink lies in the lack of knowledge on the interference from/to a large number of terrestrial transmissions in the region $D_u$ but outside the local ICIC region $D_t$. However, such interference knowledge is difficult to obtain at the UAV's serving BS $j_u$, unless there is a global information exchange among all the BSs in $D_u$, which is practically costly to implement. To resolve the above issue, we propose in this section a new UAV-sensing-assisted RB allocation scheme that requires only local information exchange between the UAV and its serving BS, by exploiting the UAV's high altitude and hence stronger sensing capability than its serving BS.\vspace{-3pt}

\subsection{UAV-Sensing-Assisted Downlink RB Allocation}
To maximize the UAV's achievable rate in the downlink, its serving BS $j_u$ should always transmit with the peak power $P_{\text{DL}}$ in each RB assigned to the UAV. Thus, the UAV's achievable sum-rate is mainly determined by the RBs allocated to it. Ideally, if the UAV's serving BS $j_u$ is aware of the channel state information (CSI) $F_{j_u}(n)$ and the UAV's received interference power $I_{\text{DL}}(n), \forall n \in \Omega_d$, the optimal RB allocation can be obtained as the $N_d$ RBs in $\Omega_d$ with the largest values of $\frac{F_{j_u}(n)}{\sigma^2 + I_{\text{DL}}(n)}$. Notice that for high-altitude UAVs, their channels with serving BSs are approximately frequency-flat due to the dominant LoS links; thus, $F_{j_u}(n)$'s do not vary much over $n$ and the optimal RBs can be obtained by searching the $N_d$ RBs in $\Omega_d$ with the lowest values of $I_{\text{DL}}(n)$. However, due to the lack of knowledge on each $I_{\text{DL}}(n)$ at BS $j_u$, it is practically difficult to implement this optimal RB allocation by BS $j_u$ directly. To overcome this difficulty, we propose the following UAV-sensing-assisted RB allocation.

Specifically, the UAV's serving BS $j_u$ first randomly selects $M_d > N_d$ candidate RBs from $\Omega_d$, denoted by $\Phi_d$ with $\lvert \Phi_d \rvert=M_d \le \lvert \Omega_d \rvert$, and sends their indices to the UAV. Then, the UAV performs spectrum sensing to measure its received interference power $I_{\text{DL}}(n), n \in \Phi_d$.
Next, the UAV selects the $N_d$ RBs with the lowest interference power among the $M_d$ candidate RBs in $\Phi_d$, denoted as $\bar\Phi_d$, for its downlink transmission. Finally, the UAV reports the indices of its selected RBs in $\bar\Phi_d$ to BS $j_u$, which assigns these RBs to the UAV's downlink transmission. As such, the proposed RB allocation takes into account all interfering BSs in $D_u$, as compared to the local region $N_{j_u}(q)$ only in the conventional ICIC. Accordingly, with the proposed ICIC, the UAV's achievable sum-rate is obtained with ${\cal N}_{\text{DL}}\!=\!\bar\Phi_d$ and $P_{j_u}(n)\!=\!P_{\text{DL}}, \forall n \in \bar\Phi_d$ in (\ref{sum1}).

\begin{remark}\label{tradeoff1}
If BS $j_u$ assigns more candidate RBs (i.e., larger $M_d$) for the UAV to sense and select, then the UAV's achievable sum-rate can be improved in general. However, this also increases the sensing cost at the UAV.
\end{remark}\vspace{-6pt}

\subsection{UAV-Sensing-Assisted Uplink RB Allocation}
In the uplink, the UAV needs to help its serving BS $j_u$ determine the RB allocation ${\cal N}_{\text{UL}}$ and its power allocation $p_u(n), n \in {\cal N}_{\text{UL}}$ to maximize its uplink transmission rate $R_{\text{UL}}$ subject to the peak power constraints $p_u(n) \le P_{\text{UL}}, \forall n \in {\cal N}_{\text{UL}}$; while effectively controlling its worst-case (maximum) interference power $I_{\text{UL}}$ in (\ref{sum4}) to any co-channel BS below a given threshold $\Gamma_u$, which can be practically chosen to be sufficiently small (e.g., in the order of BS receiver noise power $\sigma^2$). Ideally, if the UAV is aware of its interfering channels to all occupied BSs over all available RBs, i.e., $G_j(n), \forall j \in {\cal J}(n), n \in \Omega_u$, the optimal power allocation can be obtained as $p^*_u(n)=\min\{\mathop {\min}\nolimits_{j \in {\cal J}(n)}G^{-1}_j(n)\Gamma_u,P_{\text{UL}}\}, n \in {\cal N}_{\text{UL}}$, and the optimal RB allocation should assign the UAV with the $N_u$ RBs in $\Omega_u$ having the largest $N_u$ values of $p^*_u(n)G_{j_u}(n)$.
However, in practice, the UAV can only sense the uplink transmissions from the terrestrial UEs in each RB, based on which it is difficult to estimate directly the required interfering CSI to their serving BSs for the optimal RB allocation and power control solutions. To resolve the above challenge, we propose a UAV-sensing-assisted RB allocation and power control scheme by ensuring the worst-case uplink interference from the UAV to any co-channel BS that lies outside $N_{j_u}(q)$ but still in $D_u$.
\begin{figure}[!t]
\centering
\includegraphics[width=3in]{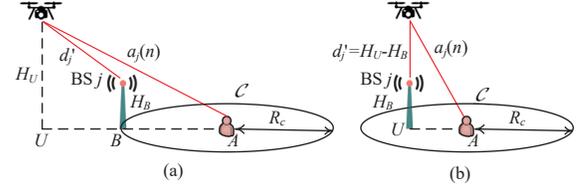}
\DeclareGraphicsExtensions.
\caption{Worst-case BS locations.}\label{geom}
\vspace{-12pt}
\end{figure}

First, similar to the downlink, the UAV's serving BS $j_u$ randomly selects $M_u > N_u$ candidate RBs from $\Omega_u$, denoted as $\Phi_u$, and sends their indices to the UAV. The UAV then performs spectrum sensing to measure its received power from all the transmitting terrestrial UEs in each candidate RB $n \in \Phi_u$, denoted as $E_{\text{UL}}(n)$. If the received power in an RB is high, it is a good indication that there may exist terrestrial UEs transmitting in the adjacent cells of the UAV (but outside $N_{j_u}(q)$), or this RB may be heavily reused by many cells in $D_u$. Thus, this RB should not be assigned to the UAV for avoiding its uplink interference to these co-channel terrestrial uplink transmissions. Accordingly, the UAV selects the $N_d$ RBs with the lowest sensed powers in $\Phi_u$, denoted as $\bar\Phi_u$, and sends them back to its serving BS for uplink transmission.
\begin{figure*}[!t]
\centering
\subfigure[]{\includegraphics[width=0.32\textwidth]{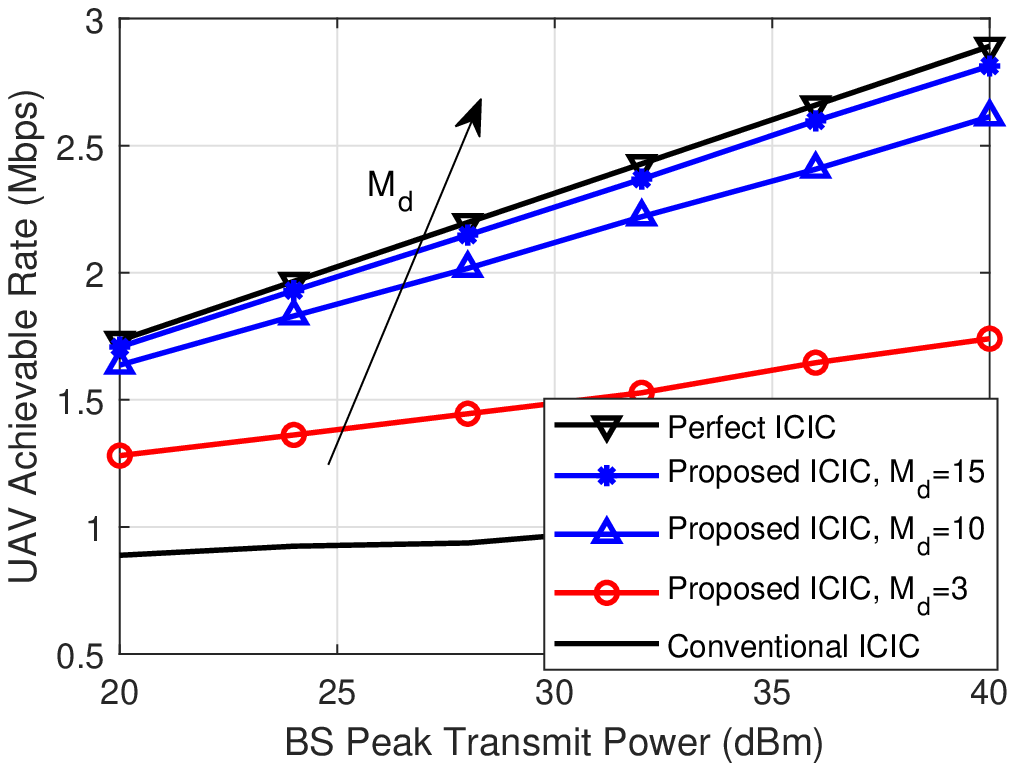}}\quad
\subfigure[]{\includegraphics[width=0.32\textwidth]{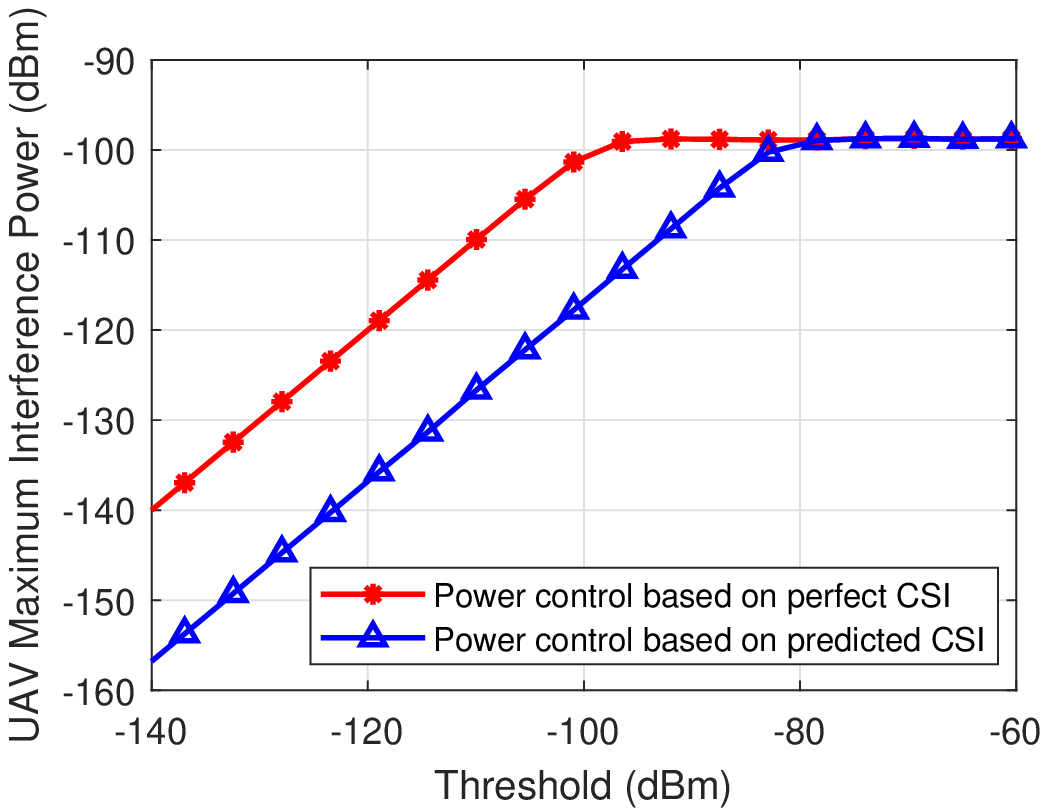}}\quad
\subfigure[]{\includegraphics[width=0.32\textwidth]{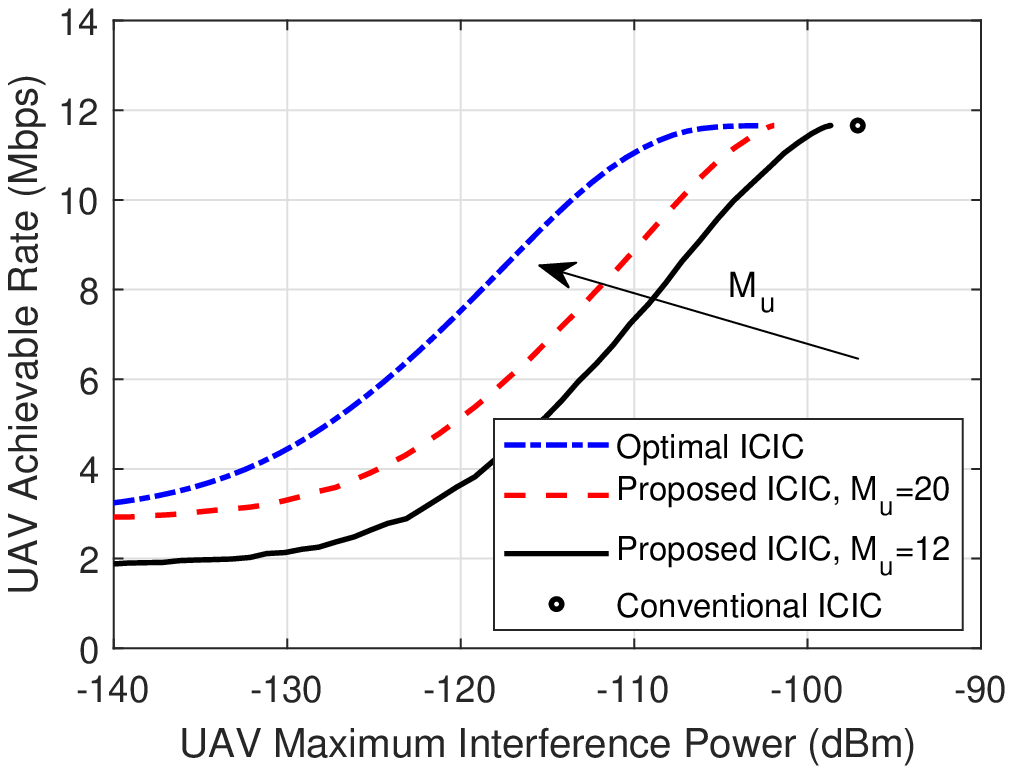}}
\caption{(a) UAV achievable rate versus BS peak transmit power in the downlink; (b) UAV maximum interference power per RB in the uplink; (c) UAV achievable rate versus maximum interference power in the uplink.}\label{sim}\vspace{-6pt}
\end{figure*}

Next, to determine the UAV's power allocation over $\bar\Phi_u$, we propose a robust power control policy by deriving an upper bound on each $G_j(n), j \in {\cal J}(n), n \in \bar\Phi_u$ based on $E_{\text{UL}}(n)$. Specifically, we first consider the worst-case interference channel power from the UAV to each BS $j \in {\cal J}(n), n \in \bar\Phi_u$, assuming the LoS channel model. Thus, we have
\begin{equation}\label{ineq1}
G_j(n) \le \beta_0 d_j^{-\alpha_L}, j \in {\cal J}(n), n \in \bar\Phi_u,
\end{equation}
where $\alpha_L$ denotes the LoS path-loss exponent in the region of $D_u$, $d_j$ denotes the distance between the UAV and BS $j$, and $\beta_0$ is the LoS channel power gain at a reference distance 1 m.

On the other hand, for the uplink transmissions sensed by the UAV, we assume that they are also dominated by LoS links and each active terrestrial UE transmits with the peak power $P_{\text{UL}}$ in each RB $n \in \bar\Phi_u$. Thus, we have
\begin{align}
E_{\text{UL}}(n) &= P_{\text{UL}}\beta_0\sum\nolimits_{j \in {\cal J}(n)}a_j^{-\alpha_L}(n) \nonumber\\
&\ge P_{\text{UL}}\beta_0a_j^{-\alpha_L}(n), \forall j \in {\cal J}(n), n \in \bar\Phi_u,\label{ineq2}
\end{align}
where $a_j(n)$ denotes the distance between the UAV and the terrestrial UE served by BS $j$ in RB $n$. To relate the results in (\ref{ineq2}) to (\ref{ineq1}), we present the following lemma.
\begin{lemma}
Let $H_u$, $H_B$ and $R_c$ denote the UAV's altitude, the BS height and the cell radius, respectively, with $H_u > H_B$. If the heights of all terrestrial UEs are ignored for simplicity, it must hold that
\begin{equation}\label{ineq3}
\frac{d_j}{a_j(n)} \!\!\ge\!\! \sqrt \frac{{(\xi - R_c)}^2 + {(H_u - H_B)}^2}{\xi^2 + H_u^2} \!\triangleq \!\rho, \forall j \in {\cal J}(n), n \in \bar\Phi_u,
\end{equation}
where $\xi = \frac{R_c^2+H_B^2-2H_uH_B+ \sqrt {(R_c^2+H_B^2-2H_uH_B)^2 + 4R_c^2H_u^2}}{2R_c}$.
\end{lemma}
\begin{IEEEproof}
As shown in Fig.\,\ref{geom}, consider an active terrestrial UE in RB $n \in \bar\Phi_u$ served by BS $j \in {\cal J}(n)$, with its location denoted as $A$. Next, we derive the worst-case location of BS $j$ that leads to the shortest distance to the UAV, denoted as $d_j'$. Obviously, we have $d_j \ge d_j'$. Notice that BS $j$ must reside in a circle $\cal C$ centered at $A$ with a radius of $R_c$. Based on this fact, we consider the following two cases. First, as shown in Fig.\,\ref{geom}(a), if the UAV's horizontal location (denoted as $U$) is outside $\cal C$, i.e., $a_j^2(n) \ge R_c^2+H_u^2$, the worst-case BS location should be the point of intersection between $\cal C$ and the line segment $UA$, denoted by $B$. It is easy to verify that $\frac{{d_j'}^2}{a_j^2(n)}\!=\! \frac{(\sqrt{a_j^2(n) - H_u^2}- R_c)^2 + (H_u - H_B)^2}{a_j^2(n)}$. By taking the derivative of the above ratio with regard to $a_j(n)$, the value of $a_j(n)$ that minimizes ${d_j'}^2/a_j^2(n)$ (or equivalently, ${d_j'}/a_j(n)$) can be found, subject to the condition $a_j^2(n) \ge R_c^2+H_u^2$. Accordingly, we can obtain $d_j'/a_j(n) \ge \rho$. On the other hand, as shown in Fig.\,\ref{geom}(b), if $U$ is inside $\cal C$, i.e., $a_j^2(n) \le R_c^2+H_u^2$, then the worst-case BS location should be $U$. Thus, we have $d_j'=H_u-H_B$ and $\frac{d_j'}{a_j(n)} \ge \frac{H_u-H_B}{\sqrt{R_c^2+H_u^2}}>\rho$. By combining the above two cases, it follows that $d_j/a_j(n) \ge d_j'/a_j(n) \ge \rho$, i.e., (\ref{ineq3}) holds. The proof is thus completed.
\end{IEEEproof}

By substituting (\ref{ineq3}) into (\ref{ineq2}), we have $\beta_0d_j^{-\alpha_L}\!<\!\rho^{-\alpha_L}\frac{E_{\text{UL}}(n)}{P_{\text{UL}}}$. From (\ref{ineq1}), it is obvious that an upper bound on each $G_j(n), j \in {\cal J}(n), n \in \bar\Phi_u$ can be obtained as $\tilde G(n) \triangleq \rho^{-\alpha_L}\frac{E_{\text{UL}}(n)}{P_{\text{UL}}}, n \in \bar\Phi_u$, which is proportional to the UAV sensed power $E_{\text{UL}}(n)$. Assuming that $\rho$ and $\alpha_L$ are known {\it a priori} at the UAV, its uplink transmit power allocations are thus given by $\bar p_u(n)=\min\{1,\frac{\Gamma_u\rho^{\alpha_L}}{E_{\text{UL}}(n)}\}P_{\text{UL}}, n \in \bar\Phi_u$. Hence, if the proposed RB allocation and robust power control scheme is applied, we set ${\cal N}_{\text{UL}}=\bar\Phi_u$ and $p_u(n)=\bar p_u(n), \forall n \in \bar\Phi_u$ in both (\ref{sum3}) and (\ref{sum4}), to obtain the UAV's achievable sum-rate and its worst-case interference power per RB to any co-channel BS, respectively.
\begin{remark}\label{tradeoff2}
If the UAV's serving BS $j_u$ assigns more candidate RBs (i.e., larger $M_u$) for the UAV to sense and select, then the tradeoff between the UAV's achievable sum-rate and its maximum interference power to any co-channel BS can be better reconciled. This is because the UAV is more likely to find RBs with low sensed power levels, such that more transmit power can be allocated to these RBs, thus improving the former but without increasing the latter. However, this comes at the expense of higher sensing cost at the UAV.
\end{remark}

\section{Numerical Results}
In this section, numerical results are provided to compare the performance of the proposed UAV-sensing-assisted ICIC with that of the conventional ICIC presented in Section \ref{conv}, as well as the optimal ICIC which applies the optimal RB allocation and power control, assuming that all the needed CSI is perfectly known. Unless otherwise specified, the simulation settings are as follows. The tier of neighboring BSs is $q=1$ in the conventional ICIC for both downlink and uplink. The total number of RBs is $N=30$, each with a bandwidth of $B=180$ kHz. The carrier frequency $f_c$ is $2$ GHz, and the noise power spectrum density at the UAV/BS receiver is $-164$ dBm/Hz. The cell radius is $R_c=800$ m, and the height of BSs is set to be $H_B = 25$ m. The altitude of the UAV is fixed as $H_u=200$ m. The BS antenna elements are placed vertically with half-wavelength spacing and electrically steered with 10-degree downtilt angle. We consider three tiers of cells centered at the cell underneath the UAV (named cell 1) to cover $D_u$, and thus the total number of cells is $J=37$. The total number of terrestrial UEs over the $N$ RBs in $D_u$ is $60$. The UAV's horizontal location is randomly generated in cell 1, while the terrestrial UEs' locations are randomly generated in the $J$ cells. The path-loss of all BS-UAV channels follows the probabilistic LoS channel model in the urban macro scenario in \cite{3GPP36777}, while their small-scale fading is modeled as Rician fading with a Rician factor of $20$ dB. Note that with $H_u = 200$ m, we have $\beta_0 = -34$ dB and $\alpha_L = 2.2$\cite{3GPP36777}. The number of requested RBs by the UAV is set to be $N_d=1$ and $N_u=10$ for its downlink and uplink communications, respectively. All results shown are averaged over 1000 random channel and location realizations of the terrestrial UEs and the UAV.

First, in Fig.\,\ref{sim}(a), we show the UAV's achievable rates in the downlink, $R_{\text{DL}}$, by all considered ICIC designs versus the BS peak transmit power, $P_{\text{DL}}$. It is observed that the proposed ICIC significantly outperforms the conventional ICIC, thanks to the enlarged interference mitigation region ($D_u$ versus $D_t$). Moreover, in accordance with Remark \ref{tradeoff1}, it is observed that the UAV's achievable rate by the proposed ICIC increases with $M_d$. In particular, the performance gap between the proposed ICIC and the optimal ICIC becomes negligible when $M_d\!\ge\!15$.

Next, in Fig.\,\ref{sim}(b), we show the UAV's maximum interference power per RB in $\bar\Phi_u$ to any co-channel BS in the uplink, i.e., $I_{\text{UL}}$ given in (\ref{sum4}), by the proposed power control based on the predicted CSI (i.e., $\tilde G(n), n \in \bar\Phi_u$) versus the interference power threshold $\Gamma_u$, with $P_{\text{UL}}=10$ dBm and $M_u=12$. We also plot the result by the power control based on the perfect CSI, assuming that each $G_j(n), j \in {\cal J}(n), n \in \bar\Phi_u$ is known at the UAV. From Fig.\,\ref{sim}(b), it is observed that with the proposed power control, the UAV's maximum interference power is ensured to be no greater than the given threshold $\Gamma_u$. However, due to our worst-case analysis, there exists a gap (or over-protection for co-channel BSs) between them.

Last, Fig.\,\ref{sim}(c) plots the UAV's achievable rate in the uplink, $R_{\text{UL}}$, under different ICIC designs versus $I_{\text{UL}}$, by varying the interference threshold $\Gamma_u$. The uplink peak transmit power is set to be $P_{\text{UL}}=10$ dBm. It is observed that the considered ICIC designs have almost the same UAV's maximum achievable rate, when $I_{\text{UL}}$ becomes sufficiently large. However, the proposed ICIC yields a more flexible tradeoff between $R_{\text{UL}}$ and $I_{\text{UL}}$, as well as lower $I_{\text{UL}}$ than the conventional ICIC. It is also observed that increasing $M_u$ helps improve this tradeoff, which is in accordance with Remark \ref{tradeoff2}.

\section{Conclusions}
This letter proposes a new cognitive radio approach for resolving the challenging interference issue for cellular-connected UAVs. It exploits the UAV's high altitude and resultant LoS-dominant channels with ground BSs/UEs to perform efficient spectrum sensing to assist its serving BS in RB allocation and power control, which can be practically implemented in a low-complexity and distributed manner. Numerical results show that the proposed ICIC significantly improves the UAV's achievable rate in the downlink as compared to the conventional ICIC. Whereas in the uplink, it enables a more flexible tradeoff between the UAV's achievable rate and maximum interference power to any co-channel terrestrial transmission. This letter can be extended in several directions in future work, such as multi-antenna UAVs, massive multi-input multi-output (MIMO) BSs, and/or muti-UAV cooperative sensing and RB allocation.

\bibliography{UAV_SS}
\bibliographystyle{IEEEtran}

\end{document}